\documentstyle{article}
\begin{document} 
\begin{center}
{\bf Spin glasses at imaginary temperature.}\\
{D.B. Saakian}\\
{Yerevan Physics Institute,
Alikhanian Brothers St. 2, Yerevan 375036, Armenia }
\end{center}
\begin{abstract}
We consider spherical p-spin glass and p-spin glass models at imaginary  temperatures. 
Imaginary temperatures are special case, when order parameters are real value numbers. Here
 there is a some antiferromagnetic like order.
\end{abstract}                   
\vspace{10mm}

The spin-glass models [1-2] have many applications in different  fields of physics. 
The phase structure of  mean field models at real temperatures is well known.
Last years there was a great interest to investigation of the statstical mechanics models at complex temperatures. An investigation of models for the complex temperatures reveals some additional properties of the models.The case of models at imaginary tem

peratures is not  a pure mathematical 
 exercise, there are  real physical  situations [3], which can be considered as (finite replica) SG models at at imaginary temperatures.\\
 For the case of Random 
Energy Model (REM) and Generalized REM  (GREM) the phase structure has been found
 [4-5]. Besides the paramagnetic  and spin-glass phases there is a third one: the  
 Lee-Young-Fisher phase. 
 At the consideration of complex temperatures, the complex order parameters are not often 
 convenient to manipulate. Hopefully there is an exception in the case of imaginary temperatures
 and odd p. Here the contribution of inverse spin configurations are 
 conjugate to each other and thus both the spin-spin correlator and the partition are real numbers. 
 In this work  we are going to find the phase structure of p-spherical model and p spin model 
 at imaginary  temperatures and odd values of p.\\
Let us consider first Hamiltonian 
\begin{equation}
\label{e1}
H=-\sum_{1\le i_1<i_2..<i_p\le N}j_{i_1..i_p}s_{i_1}..s_{i_p}
\end{equation}
with continuos spins $s_i$
\begin{equation}
\label{e2}
\sum_{i=1}^N s_{i}^2=N
\end{equation}
and $M=\frac{N!}{p!(N-p)!}$ couplings with mean values
\begin{equation}
\label{e3}
\overline {j_{i_1..i_p}^2}=\frac{N}{2M}
\end{equation}
We take complex temperatures with
\begin{equation}
\label{e4}
\frac{1}{T}=i\beta
\end{equation}
To calculate quenched free energy one can use replica approach. Introducing order parameter
q and Lagrange conjugate variables $\lambda$one can deduce  as in [6]:
\begin{eqnarray}
\label{e5}
\overline {Z^n}=\int \prod_{\alpha<\beta} dq_{\alpha\beta}\int _{-i\infty}^{i\infty}
\prod_{\alpha<\beta}\frac{N}{2\pi i}d\lambda_{\alpha\beta}\int _{-i\infty}^{i\infty}
\prod_{\alpha}\frac{N}{4\pi i}d\lambda_{\alpha\alpha}\exp[NG]\nonumber\\
G=-\frac{\beta^2}{4}\sum_{\alpha<\beta}q_{\alpha\beta}^p-\frac{1}{2}\sum_{\alpha<\beta} q_{\alpha\beta}{\lambda}_{\alpha\beta}
+\ln \int_{-i\infty}^{i\infty}\prod_{\alpha}d\sigma_{\alpha}\exp[\sum_{\alpha<\beta}
\frac{1}{2}\lambda_{\alpha\beta}
\sigma_{\alpha}\sigma_{\beta}]
\end{eqnarray}
We see, that this expression resembles the case of real temperatures [6]. For the 
paramagnetic phase one can take  the same expressions (6)-(9) as in [6], only we should replace 
$\beta^2\to -\beta^2$.\\
Integration via $d\sigma$ gives :
\begin{eqnarray}
\label{e6}
\frac{n}{2}\ln 2\pi-\frac{1}{2}\ln det\{-\lambda_{\alpha\beta}\}
\end{eqnarray}
Variation via q gives
\begin{eqnarray}
\label{e7}
\lambda_{\alpha\beta}^{-1}=q_{\alpha\beta}
\end{eqnarray}
\begin{eqnarray}
\label{e8}
G=\frac{\beta^2}{4}\sum_{\alpha\beta}q_{\alpha\beta}^p+\frac{n(1+\ln 2\pi)}{2}
+\frac{1}{2}Tr\ln q_{\alpha\beta}
\end{eqnarray}
For the paramagnetic phase we take  $q_{\alpha\beta}=0$ for $\alpha\ne\beta$
 and thus we have for PM solution an expression:
\begin{eqnarray}
\label{e9}
\frac{<ln Z>}{N}=-\frac{\beta^2}{4}+\frac{(1+\ln 2\pi)}{2}
\end{eqnarray}
Let us consider now the second, Lee-Young-Fisher phase [7-8] for the
case of odd p. Here the partition is a real number, as well as $q_{\alpha\beta}$.\\
\begin{eqnarray}
\label{e10}
-\frac{\beta^2(1+\sum_{\alpha\ne\beta} q_{\alpha\beta}^p}{4})+
\frac{1}{2}\sum_{\alpha\beta}\lambda_{\alpha\beta}
q_{\alpha\beta}+
\ln \int_{-i\infty}^{i\infty}\prod_{\alpha}d\sigma_{\alpha}\exp[\sum_{\alpha\beta}
\frac{1}{2}\lambda_{\alpha\beta}
\sigma_{\alpha}\sigma_{\beta}]
\end{eqnarray}
We divide  all the replicas into two classes. The nonzero elements of order parameter q are
 on diagonal $q_{\alpha\alpha}=1$ and $q_{\alpha,\alpha+n/2}=q,1\le n/2$.
 We have for the free energy:
\begin{eqnarray}
\label{e11}
\ln \frac{<Z^n>}{N}=-\frac{\beta^2n}{4}[1+q^p]+\frac{\lambda_0n}{2}+\frac{q\lambda_1n}{2}+\nonumber\\
+\ln\int\exp[-\sum_{\alpha=1}^{n/2}\frac{1}{2}\lambda_0(x_{\alpha}^2+y_{\alpha}^2)
+\lambda_1x_{\alpha}y_{\alpha}]=\nonumber\\
=-\frac{\beta^2n}{4}[1+q^p]+\frac{\lambda_0n}{2}+\frac{q\lambda_1n}{2}+
+\frac{n}{2}\ln\int\exp[-\frac{1}{2}\lambda_0(x^2+y^2)
+\lambda_1xy]\nonumber\\
=-\frac{\beta^2n}{4}[1+q^p]+\frac{\lambda_0n}{2}+\frac{q\lambda_1n}{2}+\nonumber\\
\frac{n}{2}\ln\int\exp[-\frac{1}{4}[(\lambda_0+\lambda_1)(x+y)^2+(\lambda_0-\lambda_1)(x-y)^2]\nonumber\\
-\frac{\beta^2n}{4}[1+q^p]+\frac{\lambda_0n}{2}+\frac{q\lambda_1n}{2}+
n/2\ln 2\pi-n/4\ln (\lambda_0^2-\lambda_1^2) 
\end{eqnarray}
We have the saddle point equations:
\begin{eqnarray}
\label{e12}
\frac{\beta^2p}{4}q^{p-1}=\lambda_1\nonumber\\
\lambda_0=\lambda_0^2-\lambda_1^2\nonumber\\
(\lambda_0^2-\lambda_1^2)q=-\lambda_1
\end{eqnarray}
Putting the values q we obtain for the $\lambda_0,\lambda_1$:
\begin{eqnarray}
\label{e13}
\lambda_0-\lambda_0^2+(\lambda_0)^{\frac{2(p-1)}{p-2}}(\frac{4}{\beta^2p})^{\frac{2}{p-2}}=0\nonumber\\
(\lambda_0)^{\frac{p-1}{p-2}}(\frac{4}{\beta^2p})^{\frac{1}{p-2}}=\lambda_1
\end{eqnarray}
Having the values of $\lambda$ we can calculate the main order parameter:
\begin{equation}
\label{e14}
q=-(\frac{4}{p\beta^2}\lambda_0)^{\frac{1}{p-2}}
\end{equation}
Now we can calculate also the free energy:
\begin{equation}
\label{e15}
\ln \frac{<Z^n>}{N}=-\frac{\beta^2n}{4}[1+q^p]+\frac{1}{2}-\frac{n}{4}\ln \lambda_0
\end{equation}
We see, that 
\begin{equation} 
\label{e16}
0>q>-1,\lim_{p\to \infty} q\to -1
\end{equation}
Another limit at $p\to \infty$:
\begin{eqnarray}
\label{e17}
\ln \frac{<Z^n>}{N}=
-1/4\ln p
\end{eqnarray}
We have transition  between two phases, when (9) coincides with the expression of free 
energy from the (15). \\
Let us consider the case of p-spin interaction (1), when the spins $s_i $ are $\pm 1$.
We have:
\begin{eqnarray}
\label{e18}
\overline{Z^n}=\int \prod_{\alpha<\beta} dq_{\alpha\beta}\int _{-i\infty}^{i\infty}
\prod_{\alpha<\beta}\frac{N}{2\pi i}d\lambda_{\alpha\beta}\int _{-i\infty}^{i\infty}
\exp[NG]\nonumber\\
G=-\frac{\beta^2}{4}[1+\sum_{\alpha<\beta}q_{\alpha\beta}^p]-\frac{1}{2}\sum_{\alpha<\beta} q_{\alpha\beta}{\lambda}_{\alpha\beta}
+\ln Tr\exp[\sum_{\alpha<\beta}
\frac{1}{2}\lambda_{\alpha\beta}
\sigma_{\alpha}\sigma_{\beta}]
\end{eqnarray} 
For paramagnetic solution $q=0,\lambda=0$ one has a trivial expression:
\begin{eqnarray}
\label{e19}
\frac{<\ln Z>}{N}=-\frac{\beta^2}{4}+\ln 2
\end{eqnarray} 
For the second phase we take again fracture n replicas into two classes, and use an anzats
 for nonzero components $q_{\alpha,\alpha+n/2}=q,\lambda_{\alpha,\alpha+n/2}=\lambda,1\le n/2$.			 
\begin{eqnarray}
\label{e20}
\overline{Z^n}=\int \prod_{\alpha<\beta} dq_{\alpha\beta}\int _{-i\infty}^{i\infty}
\prod_{\alpha<\beta}\frac{N}{2\pi i}d\lambda_{\alpha\beta}\int _{-i\infty}^{i\infty}
\exp[NGn]\nonumber\\
G=-\frac{\beta^2}{4}[1+q^p]+\frac{1}{2}q\lambda
+\frac{1}{2}\ln Tr\exp[\sigma_1\sigma_2]=\nonumber\\
-\frac{\beta^2}{4}[1+q^p]+\frac{1}{2}q\lambda
+\frac{1}{2}\ln \cosh \lambda+\ln 2 
\end{eqnarray} 
We have:
\begin{eqnarray}
\label{e21}
\frac{p\beta^2(\tanh \lambda)^{p-1}}{2}=\lambda\nonumber\\
\frac{<\ln Z>}{N}=
-\frac{\beta^2}{4}[1+q^p]+\frac{1}{2}\lambda\tanh \lambda
+\frac{1}{2}\ln \cosh \lambda+\ln 2
\end{eqnarray}
\begin{equation}
\label{e22}
q=-\tanh \lambda
\end{equation} 
We see, that at the limit $p\to \infty$
\begin{eqnarray}
\label{e23}
q\to -1,
\frac{<\ln Z>}{N}=-\frac{1}{2}\ln 2
\end{eqnarray} 

We investigated mean-field spin glass models (p-spin and p-spherical
 spin models) with p-spin interactions in case for odd values
of p. This a special situation of complex temperatures, when replica symmetry parameters are 
real numbers. This is the first investigation of Lee-Young-Fisher phase in Parisi scheme of replica symmetery breaking. There is an antiferromagnetic like order (negative value of q in Eqs.
(14),(16) and (22),(23)). A simple finite 
 replica symmetry breaking has been established. We did not give a rigorous prove of the 
 stability of our solution, but there are no any physical reasons for the existence of SG phase 
 and fore more serious breaking of replica symmetry.\\
The case of odd p is more complicated, but again there should be some sort of antiferromagnetic 
ordering. The results of SK model for replica symmetry breaking are relevant also for SG models
in finite dimensions. We hope, that the same should be the situation with weak antiferromagnetic order found in this work.

This work was supported by ISTC fund Grant A-102.

\end{document}